\documentclass[twocolumn]{aastex63}

\shorttitle{A Luminous Quasar at $z=7.642$}
\shortauthors{Wang et al.}

\begin{document}

\title{A Luminous Quasar at Redshift 7.642}

\correspondingauthor{Feige Wang}
\email{feigewang@email.arizona.edu}

\author[0000-0002-7633-431X]{Feige Wang}
\altaffiliation{Hubble Fellow}
\affil{Steward Observatory, University of Arizona, 933 North Cherry Avenue, Tucson, AZ 85721, USA}

\author[0000-0001-5287-4242]{Jinyi Yang}
\altaffiliation{Strittmatter Fellow}
\affil{Steward Observatory, University of Arizona, 933 North Cherry Avenue, Tucson, AZ 85721, USA}

\author[0000-0003-3310-0131]{Xiaohui Fan}
\affil{Steward Observatory, University of Arizona, 933 North Cherry Avenue, Tucson, AZ 85721, USA}

\author[0000-0002-7054-4332]{Joseph F. Hennawi}
\affil{Department of Physics, University of California, Santa Barbara, CA 93106-9530, USA}
\affil{Max Planck Institut f\"ur Astronomie, K\"onigstuhl 17, D-69117, Heidelberg, Germany}

\author[0000-0002-3026-0562]{Aaron J. Barth}
\affil{Department of Physics and Astronomy, University of California, Irvine, CA 92697, USA}

\author[0000-0002-2931-7824]{Eduardo Banados}
\affil{Max Planck Institut f\"ur Astronomie, K\"onigstuhl 17, D-69117, Heidelberg, Germany}

\author[0000-0002-1620-0897]{Fuyan Bian}
\affil{European Southern Observatory, Alonso de C\'ordova 3107, Casilla 19001, Vitacura, Santiago 19, Chile}

\author[0000-0003-4432-5037]{Konstantina Boutsia}
\affil{Las Campanas Observatory, Carnegie Observatories, Colina El Pino, Casilla 601, La Serena, Chile}

\author[0000-0002-7898-7664]{Thomas Connor}
\affiliation{Jet Propulsion Laboratory, California Institute of Technology, 4800 Oak Grove Drive, Pasadena, CA 91109, USA}

\author[0000-0003-0821-3644]{Frederick B. Davies}
\affil{Lawrence Berkeley National Laboratory, 1 Cyclotron Rd, Berkeley, CA 94720, USA}
\affil{Max Planck Institut f\"ur Astronomie, K\"onigstuhl 17, D-69117, Heidelberg, Germany}

\author[0000-0002-2662-8803]{Roberto Decarli}
\affil{INAF -- Osservatorio di Astrofisica e Scienza dello Spazio di Bologna, via Gobetti 93/3, I-40129 Bologna, Italy}

\author[0000-0003-2895-6218]{Anna-Christina Eilers}
\altaffiliation{Hubble Fellow}
\affil{MIT-Kavli Institute for Astrophysics and Space Research, 77 Massachusetts Avenue, Building 37, Room 664L, Cambridge, Massachusetts 02139, USA}

\author[0000-0002-6822-2254]{Emanuele Paolo Farina}
\affiliation{Max Planck Institut f\"ur Astrophysik, Karl--Schwarzschild--Stra{\ss}e 1, D-85748, Garching bei M\"unchen, Germany}

\author[0000-0003-1245-5232]{Richard Green}
\affil{Steward Observatory, University of Arizona, 933 North Cherry Avenue, Tucson, AZ 85721, USA}

\author[0000-0003-4176-6486]{Linhua Jiang}
\affil{Kavli Institute for Astronomy and Astrophysics, Peking University, Beijing 100871, China}

\author[0000-0001-6239-3821]{Jiang-Tao Li}
\affil{Department of Astronomy, University of Michigan, 311 West Hall, 1085 S. University Ave, Ann Arbor, MI, 48109-1107, USA}

\author[0000-0002-5941-5214]{Chiara Mazzucchelli}
\affil{European Southern Observatory, Alonso de C\'ordova 3107, Casilla 19001, Vitacura, Santiago 19, Chile}

\author[0000-0002-2579-4789]{Riccardo Nanni}
\affil{Department of Physics, University of California, Santa Barbara, CA 93106-9530, USA}

\author[0000-0002-4544-8242]{Jan-Torge Schindler}
\affil{Max Planck Institut f\"ur Astronomie, K\"onigstuhl 17, D-69117, Heidelberg, Germany}

\author[0000-0001-9024-8322]{Bram Venemans}
\affil{Max Planck Institut f\"ur Astronomie, K\"onigstuhl 17, D-69117, Heidelberg, Germany}

\author[0000-0003-4793-7880]{Fabian Walter}
\affil{Max Planck Institut f\"ur Astronomie, K\"onigstuhl 17, D-69117, Heidelberg, Germany}

\author[0000-0002-7350-6913]{Xue-Bing Wu}
\affil{Kavli Institute for Astronomy and Astrophysics, Peking University, Beijing 100871, China}
\affil{Department of Astronomy, School of Physics, Peking University, Beijing 100871, China}

\author[0000-0002-5367-8021]{Minghao Yue}
\affil{Steward Observatory, University of Arizona, 933 North Cherry Avenue, Tucson, AZ 85721, USA}

\begin{abstract}
Distant quasars are unique tracers to study the formation of the earliest supermassive black holes (SMBHs) and the history of cosmic reionization. Despite extensive efforts, only two quasars have been found at $z\ge7.5$, due to a combination of their low spatial density and the high contamination rate in quasar selection. We report the discovery of a luminous quasar at $z=7.642$, J0313$-$1806, the most distant quasar yet known. This quasar has a bolometric luminosity of $3.6\times10^{13} L_\odot$. Deep spectroscopic observations reveal a SMBH with a mass of $(1.6\pm0.4) \times10^9M_\odot$ in this quasar. The existence of such a massive SMBH just $\sim$670 million years after the Big Bang challenges significantly theoretical models of SMBH growth. In addition, the quasar spectrum exhibits strong broad absorption line (BAL) features in \ion{C}{4} and \ion{Si}{4}, with a maximum velocity close to 20\% of the speed of light. The relativistic BAL features, combined with a strongly blueshifted \ion{C}{4} emission line, indicate that there is a strong active galactic nucleus (AGN) driven outflow in this system. ALMA observations detect the dust continuum and [\ion{C}{2}] emission from the quasar host galaxy, yielding an accurate redshift of $7.6423 \pm 0.0013$ and suggesting that the quasar is hosted by an intensely star-forming galaxy, with a star formation rate of $\rm\sim 200 ~M_\odot ~yr^{-1}$ and a dust mass of $\sim7\times10^7~M_\odot$. Followup observations of this reionization-era BAL quasar will provide a powerful probe of the effects of AGN feedback on the growth of the earliest massive galaxies.
\end{abstract}


\section{Introduction} \label{sec_intro}
Luminous high-redshift quasars are key probes of the history of cosmic reionization. Deep spectroscopy of $z>6$ quasars indicates that the intergalactic medium (IGM) is significantly neutral at $z\gtrsim7$ \citep[e.g.][]{Banados18, Davies18a, Wang20a, Yang20a} but highly ionized at $z\lesssim6$ \citep[e.g.][]{Yang20b}, 
providing crucial information to map the cosmic reionization history in a way complementary to the cosmic microwave background (CMB) integral constraint on reionization from measurements of the electron scattering optical depth \citep[e.g.][]{Planck20}.

In addition,  the earliest supermassive black holes (SMBHs), the engines of the most distant quasars, are crucial for understanding the formation mechanisms of the first generation black hole seeds \citep[see][for a recent review]{Inayoshi20}. 
A billion solar mass SMBH at $z\sim7$, having grown at the Eddington limit since formation, requires a seed black hole of mass $\sim1000~M_\odot$ at the time the first luminous object formed in the Universe \citep[i.e., $z\sim30$,][]{Tegmark97}. This growth corresponds to a factor of $\sim10^6$ increase in mass within a mere $\sim650$ Myr.
The recent discovery of a $1.5\times10^9 M_\odot$ SMBH in a luminous quasar at $z=7.52$ poses the most stringent constraints yet on the masses of the seed black holes \citep{Yang20a}. 

After a decade of industrious searches, a sample of more than 50 quasars now exists at $6.5<z<7$ \citep[e.g.][]{Mazzucchelli17,Wang19,Yang19,Reed19,Matsuoka19a}, enabling the first statistical studies of the early accreting SMBHs deep into the epoch of reionization \citep{Wang19}. 
However, the sample of quasars at $z>7$ is still very limited, since the Ly$\alpha$ emission is redshifted to near-infrared wavelengths,  making both imaging and spectroscopic observations more challenging. In addition, the number density of $z>7$ quasars is low \citep{Wang19} while the contaminants, Galactic cool dwarfs and early type galaxies, are far more numerous. To date, there are only seven quasars known at $z>7$ \citep{Mortlock11,Banados18,Wang18, Yang19, Yang20a, Matsuoka19a,Matsuoka19b} with two of them at $z=7.5$ \citep{Banados18,Yang20a}.

\begin{figure*}[tbh]
\centering
\includegraphics[width=1.0\textwidth]{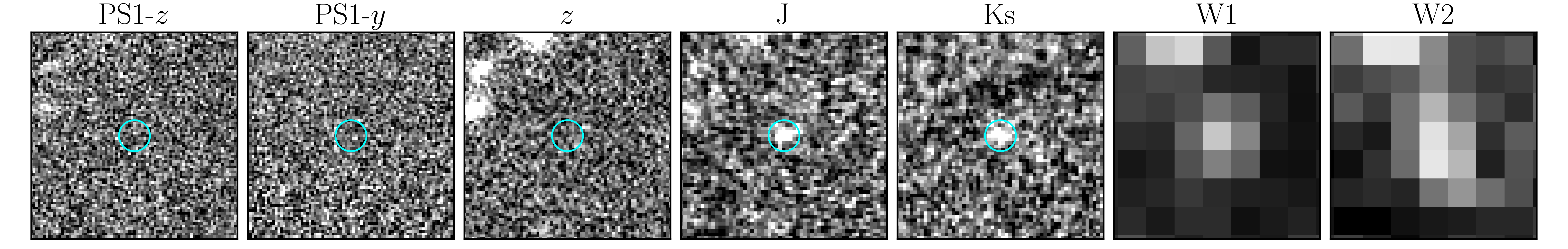}
\includegraphics[width=1.0\textwidth]{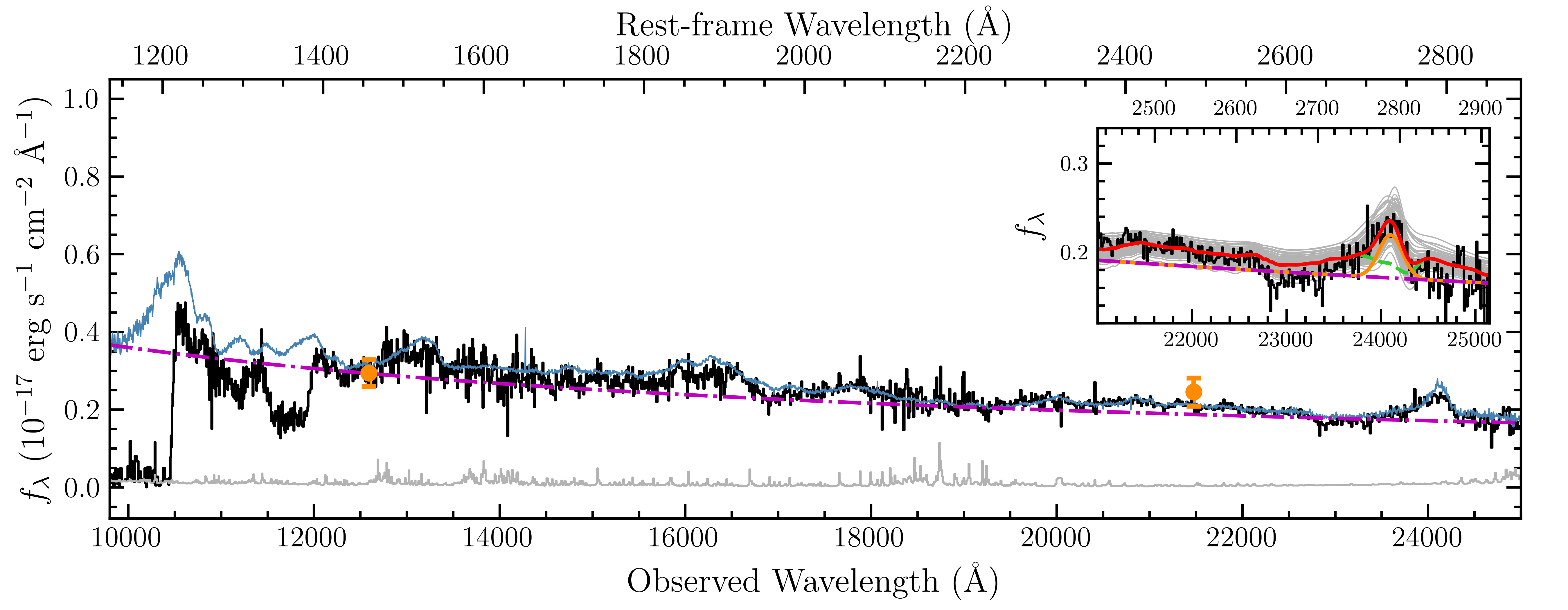}
\caption{
Upper panel: image cutouts ($20^{\prime\prime} \times 20^{\prime\prime}$, north is up and east is to the left) for J0313--1806 in PS1 $z$, PS1 $y$, DELS $z$, VISTA $J$, VISTA $K{\rm s}$, WISE W1 and WISE W2 bands. The photometry is given in Table \ref{tbl}.
Lower panel: the final stacked spectrum of J0313--1806. In the figure, we re-binned the spectrum by two spectral pixels ($\sim 173~ {\rm km~ s^{-1}}$) for illustration purposes. 
The black and gray lines represent the Galactic extinction-corrected spectrum and the error vector, respectively. The blue line denotes the quasar composite spectrum constructed with SDSS $z\sim2$ quasars having similar \ion{C}{4} blueshifts and line strengths. The purple dashed line denotes the power-law continuum.  The orange points are flux densities determined from photometry in the $J$ and $K$s-bands. The inset panel shows the \ion{Mg}{2} line fitting with the purple dot-dashed line denoting the power-law continuum, the green dashed line denoting the pseudo-continuum model (the sum of power law continuum, \ion{Fe}{2} emission, and Balmer continuum), the orange line representing the Gaussian fitting of the \ion{Mg}{2} line and the red line representing the total fit of pseudo-continuum and \ion{Mg}{2} line. The thin grey lines in the insert panel represent the spectral fitting of 100 mock spectra as described in \S \ref{sec_bh}. 
\label{fig_spec}}
\end{figure*}

In this Letter, we report the discovery of a new quasar J031343.84$-$180636.4 (hereinafter J0313--1806) at $z=7.642$, the most distant quasar known to date. The existence of a $1.6\times10^9~M_\odot$ SMBH and strong outflows in this quasar provide new insights into the formation and the growth of the earliest SMBH. Throughout this Letter, we use a $\rm\Lambda CDM$ cosmological model with $H_0=70.0~{\rm km~s^{-1}~Mpc^{-1}}$, $\rm \Omega_M=0.3$, and $\Omega_\Lambda=0.7$ and the AB photometric system.

\section{Observations and Data Reduction} \label{sec_obs}
J0313--1806 was selected as a $z>7.2$ quasar candidate from our ongoing reionization-era quasar survey \citep{Wang17,Wang18,Wang19,Yang19,Yang20a}. This survey relies on an imaging dataset that combines optical imaging from Pan-STARRS1 \citep[PS1,][]{PS1} and the DESI Legacy Imaging Surveys \citep[DELS,][]{DELS}, as well as infrared imaging from the UKIRT Hemisphere Survey \citep[UHS,][]{UHS},  the VISTA Hemisphere Survey \citep[VHS,][]{VHS}, and the {\em Wide-field Infrared Survey Explorer} survey \citep[\emph{WISE,}][]{WISE}. J0313--1806 falls into a sky area covered by PS1, DELS, VHS, and \emph{WISE}. It drops out in all of the optical bands but is well detected in the infrared bands (see Figure. \ref{fig_spec}) with colors of $z-J>3.7$, $J-{\rm W1}=0.91$, and ${\rm W1}-{\rm W2}=-0.21$. The sharp dropout in the $z$-band and the blue slope of the continuum make it a promising quasar candidate at $z>7.2$. The detailed photometry for J0313--1806 is listed in Table \ref{tbl} and the cutout images are shown in Figure \ref{fig_spec}. 

\subsection{Near-infrared Spectroscopy}
The initial spectroscopic observation for J0313--1806 was obtained on 2019 November 04 (UT) with Magellan/FIRE \citep{fire} using the high-throughput longslit mode. A 15-minute exposure shows a clear Lyman break at 1.048 $\mu$m, indicating that it is a source at $z\gtrsim7.6$. We then followed  this object up with Magellan/FIRE in Echelle mode and with the JH grism on Gemini/Flamingos-2 \citep{fl2} and confirmed it as a high redshift quasar. Since the \ion{Mg}{2} emission line, the most reliable line for measuring black hole mass at high redshifts, is redshifted to the edge of the ground-based infrared observation window, extensive spectroscopic observations were obtained with Magellan/FIRE (Echelle mode), Gemini/Flamingos-2 (with HK Grism), Keck/NIRES\footnote{\url{https://www2.keck.hawaii.edu/inst/nires/}} and Gemini/GNIRS \citep{gnirs2} to improve the S/N at this wavelength. Detailed information for all the observations is listed in Table \ref{tbl}.

All spectra were reduced with the spectroscopic data reduction pipeline {\tt PyPeIt}\footnote{\url{https://github.com/pypeit/PypeIt}} \citep{pypeit1,pypeit2}. 
The pipeline includes flat fielding, wavelength calibration, sky subtraction, optimal extraction, flux calibration and telluric correction. More detailed descriptions of each processing step can be found in \cite{pypeit2} and \cite{Wang20a}. 
We stacked all spectra obtained from FIRE (Echelle mode), Flamingos-2 (with the HK Grism), NIRES and GNIRS after binning them to a common wavelength grid with a pixel size of $\rm\sim86~km~s^{-1}$ (similar to the GNIRS native pixel scale).
We then scaled the stacked spectrum by carrying out synthetic photometry on the spectrum using the VISTA $J$-band filter response curve to match the $J$-band photometry for absolute flux calibration. The final spectrum after correcting for Galactic extinction based on the dust map of \cite{SFD98} and extinction law of \cite{EXT89} is shown in Figure \ref{fig_spec}.

\begin{deluxetable}{lr}
\tablecaption{The observational information and physical properties of J0313--1806}\label{tbl}
\startdata
\\
R.A. (J2000) & 03:13:43.84 \\
Decl. (J2000) & $-$18:06:36.4 \\
$J$ & 20.92$\pm$0.13 \\
$K_{\rm s}$ & 19.96$\pm$0.16 \\
W1 & 20.01$\pm$0.11 \\
W2 & 20.22$\pm$0.28 \\
$g$, $r$, $z$& $>26.17$, $>25.91$, $>24.63$\tablenotemark{a}  \\
$z_{\rm ps1}$, $y_{\rm ps1}$ & $>23.13$, $>22.07$\tablenotemark{a}  \\
\hline
$M_{1450}$ & $-$26.13$\pm$0.05 \\
$z_{\rm [CII]}$ & 7.6423$\pm$0.0013 \\
$z_{\rm MgII}$ & 7.611$\pm$0.004 \\
$z_{\rm CIV}$ & 7.523$\pm$0.009 \\
$\Delta v _{\rm MgII-[CII]}$ & 1072$\pm$140 km $\rm s^{-1}$ \\
$\Delta v _{\rm CIV-MgII}$ & 3080$\pm$332 km $\rm s^{-1}$ \\
$\rm FWHM_{MgII}$ & 3670$\pm$405 km $\rm s^{-1}$ \\
$\alpha_\lambda$ & $-$0.91$\pm$0.02 \\
$\lambda L_{\rm 3000\AA}$ & (2.7$\pm$0.1) $\times$ $10^{46}$ erg $\rm s^{-1}$ \\
$L_{\rm bol}$ & (1.4$\pm$0.1) $\times$ $10^{47}$ erg $\rm s^{-1}$ \\
$M_{\rm BH}$ & (1.6$\pm$0.4) $\times$ $10^{9}$ $M_\odot$ \\
$L_{\rm bol}/L_{\rm Edd}$ & 0.67$\pm$0.14 \\
$\rm FWHM_{[CII]}$ & 312$\pm$94 km $\rm s^{-1}$ \\
$F_{\rm [CII]}$ & 0.60$\pm$0.16 Jy km $\rm s^{-1}$ \\
$L_{\rm [CII]}$ & (0.80$\pm$0.22) $\times 10^9~L_{\odot}$ \\
$S_{\rm 228.4 GHz}$ & 0.45$\pm$0.05 mJy \\
$L_{\rm TIR}$ & (1.5$\pm$0.2) $\times 10^{12}~L_{\odot}$ \\
$M_{\rm dust}$ & $\sim 7 \times10^7~ M_\odot$\\
$\rm SFR_{[CII]}$ & 40$-$240 $M_\odot ~ {\rm yr^{-1}}$ \\
$\rm SFR_{TIR}$ & 225$\pm$25 $M_\odot ~ {\rm yr^{-1}}$\\
\hline
$t_{\rm exp, ALMA}$ & 29 min\\
$t_{\rm exp, FIRE, Longslit}$ & 15 min\\
$t_{\rm exp, Flamingos, JH}$ & 30 min\\
$t_{\rm exp, FIRE, Echelle}$ & 362 min\\
$t_{\rm exp, Flamingos, HK}$ & 184 min\\
$t_{\rm exp, GNIRS}$ & 485 min\\
$t_{\rm exp, NIRES}$ & 264 min\\
\enddata
\tablenotetext{a}{In these undetected bands, we report the $2\sigma$ limiting magnitudes measured from a $3\farcs0$ aperture centered at the quasar position.}
\end{deluxetable}


\begin{figure*}[tbh]
\centering
\includegraphics[width=1.0\textwidth]{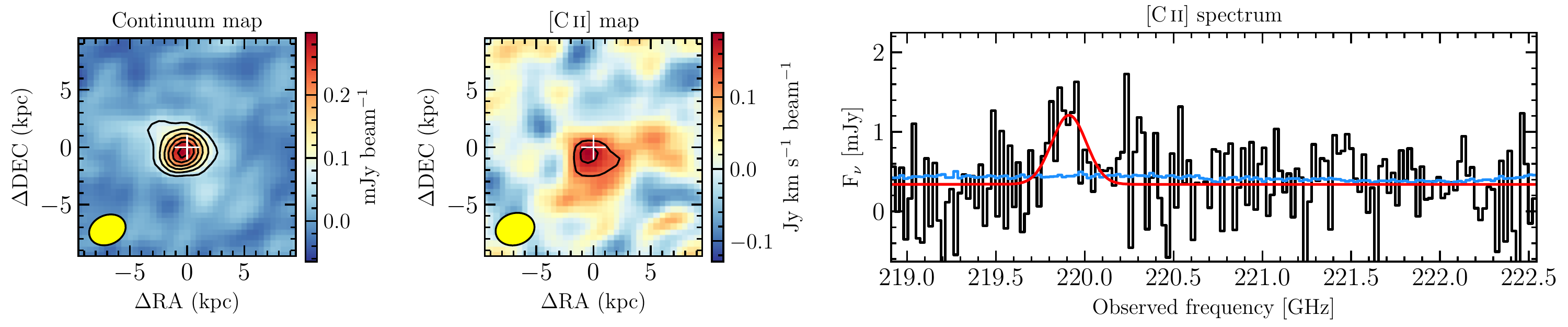}
\caption{ALMA observations of the dust continuum and [\ion{C}{2}] line. The left panel shows the dust continuum map with contour levels of [3, 5, 7, 9, 11, 13] $\times \sigma$, where $\sigma=0.02$ mJy. The middle panel shows the integrated [\ion{C}{2}] emission with contour levels of [3, 4] $\times \sigma$, where $\sigma=0.04$ Jy km $\rm s^{-1}$. The sizes of the continuum cutout and the [\ion{C}{2}] cutout are $4^{\prime\prime}\times4^{\prime\prime}$. 
The [\ion{C}{2}] intensity map was collapsed over $\pm1.4\sigma_{\rm line}$. The white crosses in both the left and middle panels highlight the quasar position derived from VHS infrared images. The right panel shows the [\ion{C}{2}] spectrum (black) and noise vector (blue) extracted from the data cube with an aperture diameter of $1\farcs5$ centered at the peak position of the continuum map. The spectrum is re-binned to 23.4 MHz channels ($\sim 30 ~{\rm km~s^{-1}}$). 
The spectral fitting (red line) gives FWHM =  $312\pm94$ km $\rm s^{-1}$ and $z_{\rm [CII]} = 7.6423\pm0.0013$.
\label{fig_c2}}
\end{figure*}

\subsection{ALMA Observations}
In order to determine the systemic redshift and investigate the host galaxy properties, we observed J0313--1806 with the Atacama Large Millimeter/submillimeter Array (ALMA) in the C43-4 configuration (program ID: 2019.A.00017.S, PI: F. Wang). We tuned two spectral windows centered at the expected frequency of the [\ion{C}{2}] line and the other two spectral windows centered at about 15 GHz away from the expected [\ion{C}{2}] line. The observations were taken in 2020 March 02 with 29 minutes of on-source integration time. 

The ALMA data were reduced using the CASA 5.6.1 pipeline \citep{CASA} following the standard calibration procedures. The continuum map was imaged by averaging over all the line-free channels using Briggs cleaning via the CASA task {\tt tclean} with robustness parameter $r = 2.0$, corresponding to natural visibility weights, to maximize the signal-to-noise ratio (S/N) of our observations. The beam size for the continuum image is $0\farcs70\times0\farcs56$. 
We subtracted the continuum using the {\tt uvcontsub} before imaging the [\ion{C}{2}] line. The [\ion{C}{2}] map was collapsed over $\pm1.4\sigma_{\rm line}$ to maximize the S/N of the intensity map.
Since both the continuum and [\ion{C}{2}] emissions are marginally resolved, we extracted the 1D spectrum with an $1\farcs$5 diameter aperture centered on the emission. The fully calibrated continuum map, [\ion{C}{2}] map and  [\ion{C}{2}]+continuum spectrum are shown in Figure \ref{fig_c2}. From the spectral fitting of the [\ion{C}{2}]+continuum spectrum, we derive a [\ion{C}{2}] redshift of $z_{\rm [CII]} = 7.6423\pm0.0013$.

\section{A 1.6 Billion Solar Mass Black Hole} \label{sec_bh}
The most reliable tool for measuring the mass of SMBHs at high redshift is the \ion{Mg}{2} virial estimator \citep{Vestergaard09}. We first de-redshift the spectrum to the rest-frame using the [\ion{C}{2}] redshift. 
Then we fit a pseudo-continuum model that contains a power-law continuum, \ion{Fe}{2} emission \citep{Vestergaard01,Tsuzuki06} and a Balmer continuum \citep{Derosa14} to spectral regions free of strong, broad emission/absorption lines (except for the \ion{Fe}{2}). 
This procedure allows us to measure a rest-frame UV slope of $\alpha_\lambda=-0.91\pm0.02$ and a quasar bolometric luminosity of $(1.4\pm0.1)\times10^{47}~{\rm erg~s^{-1}}$ after applying a bolometric correction factor of $C_{3000}=5.15$ \citep{Shen11}. 
After subtracting the pseudo-continuum model from the spectrum, we fit a two-Gaussian model to the \ion{Mg}{2} line and derive a full width at half maximum of $\rm FWHM=3670\pm405 ~km~s^{-1}$ and a \ion{Mg}{2}-based redshift of $z_{\rm MgII}=7.611\pm0.004$. 
The detailed spectral fitting result is shown in Figure \ref{fig_spec}. 
Based on the spectral fitting and the \ion{Mg}{2} virial estimator of \cite{Vestergaard09}, we estimate the mass of the central SMBH to be $M_{\rm BH}=(1.6\pm0.4)\times10^9~M_\odot$.
The Eddington ratio of this SMBH is $L_{\rm bol}/L_{\rm Edd}=0.67\pm0.14$, which indicates that the quasar is in a rapidly accreting phase, similar to other luminous quasars known at $z>7$ \citep[e.g.][]{Mortlock11,Banados18,Wang18,Yang20a}. 
Following \cite{Wang18,Wang20a} and \cite{Yang20a}, all the uncertainties are reported as the 16th and 84th percentiles of the distribution of each quantity, as derived from spectral fitting of 100 mock spectra. These mock spectra were created by randomly assigning Gaussian noise at each spectral pixel, scaled to that pixel's 1-$\sigma$ error.
All the measurements and corresponding uncertainties from the spectral fitting are listed in Table \ref{tbl}.

In Figure \ref{fig_spec}, we also show the individual fittings of the 100 mock spectra. Note that our best-fit model slightly over-estimates the flux in the ranges of 22800--23400 \AA\ and 24400--25000\AA. 
In order to explore the effects of the continuum overestimation on $M_{\rm BH}$ measurements, we re-fit the spectrum in K-band only and estimate $\rm FWHM_{MgII}=4108\pm473 ~km~s^{-1}$ and $M_{\rm BH}=(1.9\pm0.3)\times10^9~M_\odot$, consistent with that derived from the global fitting within uncertainties. 
This $K$-band spectral model gives a better fit over this wavelength range, but extrapolating it to the $J$ and $H$ bands shows that it overestimates the continuum at wavelengths shorter than $K$, 
suggesting that the quasar could be slightly reddened by dust and cannot be well modeled with a single power law.
In addition, the fitting could also be affected by the difference between the iron template and the iron emission from this particular quasar, the possible \ion{Mg}{2} absorption from outflowing gas and/or intervening metal absorbers (see \S \ref{sec_outflow}), and the large uncertainties in pixels at the red edge of the spectrum.
Future space-based observations of the spectrum beyond 2.5 $\mu$m are needed to characterize the dust extinction and to better constrain the $M_{\rm BH}$. In this work, we 
adopt the best fit from the global fitting as our fiducial model.

\begin{figure}[tbh]
\centering
\includegraphics[width=0.48\textwidth]{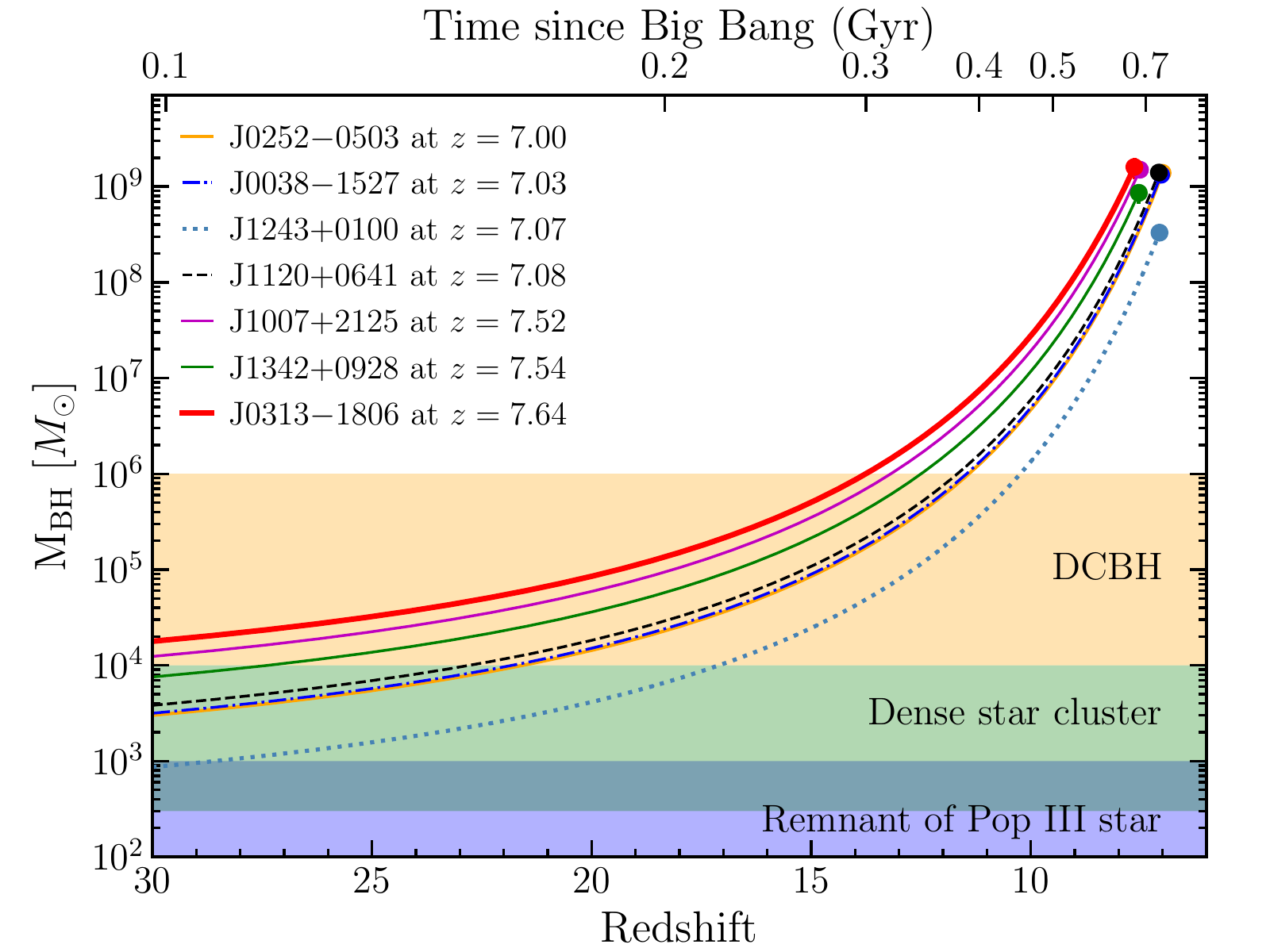}
\caption{Black hole growth track of $z\ge7$ quasars with the assumption of Eddington accretion and a radiative efficiency of 0.1 at all times. The curve gives the mass of seed black hole required to grow it to the observed mass of each SMBH in these quasars. The new quasar, J0313--0318, puts the strongest constraint on the seed black hole mass, compared to other $z>7$ quasars, including J0252--0503, J0038--1527, J1243+0100, J1120+0641, J1007+2125, and J1342+0928 \citep[see legends,][]{Yang19,Wang18,Matsuoka19b, Mortlock11,Yang20a,  Banados18}. 
The masses of the central SMBHs in the previously known quasars are collected from \cite{Matsuoka19b}, \cite{Wang18, Wang20a, Wang20b}, and \cite{Yang20a}.
The blue, green and yellow shaded regions define the rough mass ranges of the seed black hole produced by Pop III stars, dense star clusters, and direct-collapse black holes (DCBH), respectively. 
\label{fig_bh}}
\end{figure}

To compare the constraints of seed black hole masses and the growth history of the earliest SMBHs from $z>7$ quasars, we plot the growth history of all known $z>7$ quasars with $M_{\rm BH}$ measurements in Figure \ref{fig_bh} by assuming an Eddington accretion with a 10\% radiative efficiency and a duty cycle of unity. 
In this figure, all $M_{\rm BH}$ were measured using the same \ion{Mg}{2} virial estimator \citep{Vestergaard09} and thus have the same systematic uncertainties ($\sim0.5$ dex). 
J0313--1806, the highest redshift quasar known and hosting the most massive SMBH at $z>7$, poses the most challenging constraint on the seed black hole mass. 
Assuming Eddington-limited accretion, if the SMBH started its growth at redshift $z\sim15-30$ (i.e., $\sim400-570$ Myr growth time), it requires a $10^4-10^5~M_\odot$ seed black hole; such a seed is inconsistent with being a Population III star remnant \citep[e.g.][]{Madau01} or the product of dynamical processes in dense star clusters \citep[e.g.][]{Portegies04}.
Instead, direct-collapse black holes (DCBH) forming in pre-galactic dark matter halos \citep[e.g.][]{Begelman06} is the preferred seeding scenario.

\section{Extremely High Velocity Outflow} \label{sec_outflow}
\begin{figure}[tbh]
\centering
\includegraphics[width=0.48\textwidth]{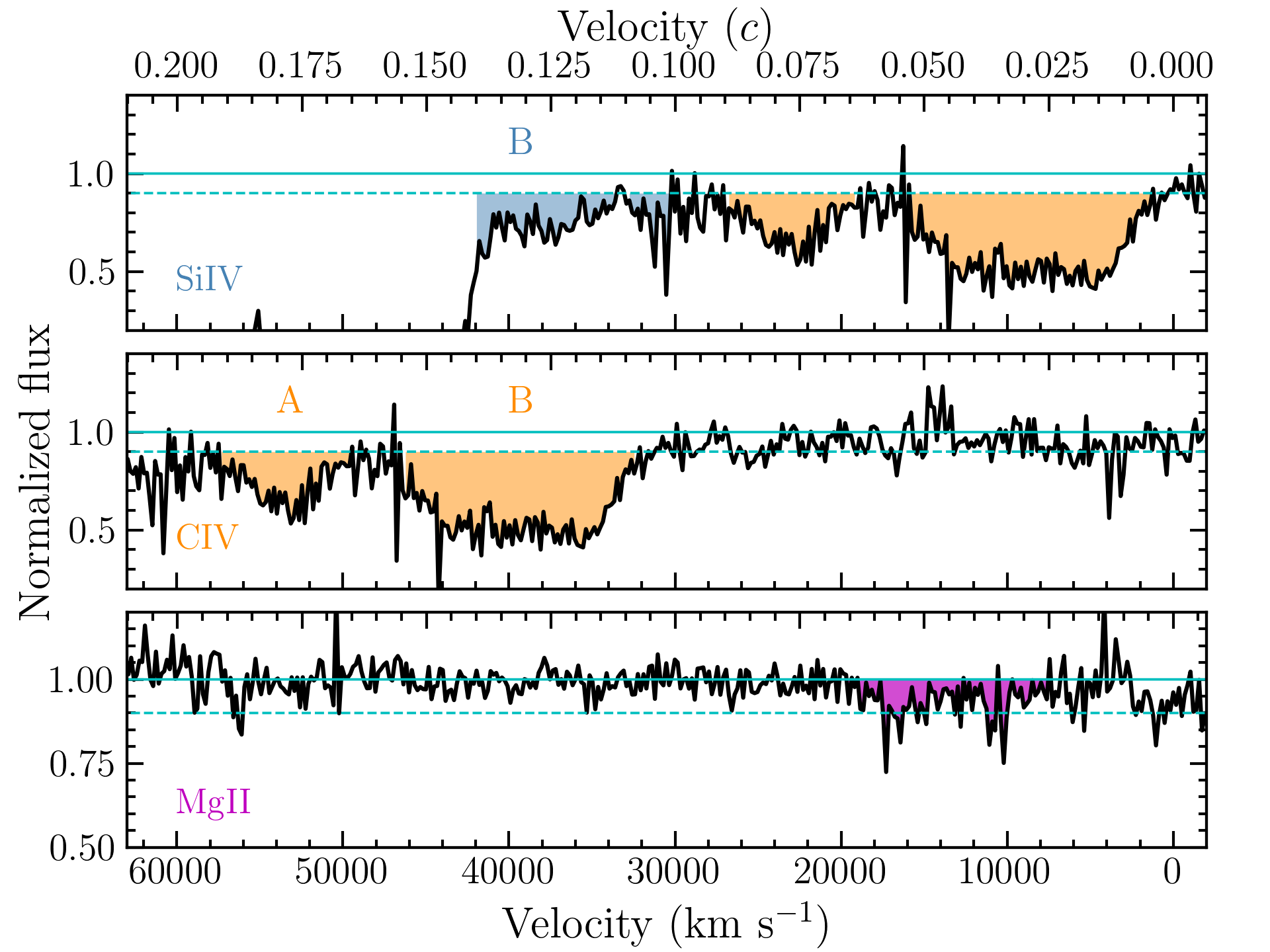}
\caption{The normalized spectrum of J0313--1806. The solid and dashed lines represent 100\% and 90\% of the normalized spectrum, respectively.
The top panel shows the outflow velocity of the \ion{Si}{4} line, the middle panel denotes the outflow velocity of the \ion{C}{4} line, and the bottom panel represents the outflow velocity of the \ion{Mg}{2} line. We interpret the two most obvious troughs (orange regions) as extremely high velocity \ion{C}{4} outflow systems with trough A having $v=(0.171-0.186)c$ and trough B having $v=(0.109-0.155)c$. 
There is a potential weak \ion{Mg}{2} absorption trough (the highlighted purple region in the bottom panel), but it does not satisfy the BI definition.
\label{fig_bal}}
\end{figure}

Another notable feature of J0313--1806 is that the spectrum contains several broad absorption line (BAL) features. The BAL features are thought to be produced by strong outflows launched from the accretion disk of the accreting SMBH. In Figure \ref{fig_bal}, we show the normalized spectra in the velocity space of the \ion{Si}{4}, \ion{C}{4}, and \ion{Mg}{2} emission lines. 
The composite spectrum (blue line in Figure \ref{fig_spec}) that was used for normalizing the observed spectrum was constructed from the spectra of a sample of low-redshift quasars with similar relative blueshifts between \ion{C}{4} and \ion{Mg}{2} lines and similar equivalent widths of \ion{C}{4} to those of J0313--1806, following the algorithm developed by \cite{Wang20a}. It was scaled to match the observed spectrum in the observed-frame $2.0-2.2~\mu$m. Using the normalized spectrum shown in Figure \ref{fig_bal}, we identified two \ion{C}{4} absorption troughs (highlighted orange regions in the middle panel of Figure \ref{fig_bal}) at extremely high velocities of (0.171--0.186)$c$ (trough A) and (0.109--0.155)$c$ (trough B). 
We use the ``balnicity'' index \citep[BI;][]{Weymann91} to estimate the strength of the BALs in the quasar. The measured BIs are 1300 km s$^{-1}$ and 5400 km s$^{-1}$ for troughs A and B, respectively. 
Trough B also has a \ion{Si}{4} absorption in the corresponding velocity range. The associated \ion{Si}{4} absorption for trough A falls into the Gunn-Peterson trough and is thus undetectable. 
Because the potential BAL absorptions from \ion{Si}{4} could clobber the proximity zone, we can not use this quasar to perform the damping wing studies. 

We also considered the alternative explanation that the two troughs are \ion{Si}{4} absorption troughs (highlighted orange regions in the top panel of Figure \ref{fig_bal}) with slightly lower velocities (i.e. $<0.1c$). We rule this explanation out given that there are no associated \ion{C}{4} absorption troughs at the corresponding velocities. We note that there is a possible weak \ion{Mg}{2} absorption (highlighted purple regions in the bottom panel of Figure \ref{fig_bal}), which would mean that J0313--1806 is a Low-ionization BAL (LoBAL). However, the \ion{Mg}{2} absorption feature does not satisfy the BI definition (i.e. continuously smaller than 0.9 for more than 2000 km s$^{-1}$). The absorption could also be affected by a mismatch of the iron emission between the composite and J0313--1806. Foreground absorption from the intervening IGM and/or circumgalactic medium (CGM) could also contribute to some of the absorption. Future high resolution spectroscopic observations are needed to identify individual foreground metal absorbers.

Extremely high velocity ($>0.1~c$) outflows are among the most promising evidence for active galactic nucleus (AGN) feedback, given that their kinetic power might be high enough to affect the star formation in quasar host galaxies \citep[e.g.][]{Chartas09}.
Such outflows are also very rare phenomena at lower redshifts; for example, 
\cite{Rodriguez20} identified only 40 ($\sim0.6$\%) extremely high velocity outflow systems from a parent sample of $\sim6700$ luminous quasars at $z\sim2-5$. Thus, the discovery of relativistic outflows in J0313--1806 and J0038--1527 \citep[$z=7.03$;][]{Wang18} among a sample of only 8 $z>7$ quasars indicates that relativistic outflows are more common at the highest redshifts, suggesting an evolution with redshift of the quasar outflow phenomenon. 
In addition, the \ion{C}{4} broad emission line has a substantial blueshift, another signature of radiation driven outflows, with a velocity of 3080 km s$^{-1}$ relative to the \ion{Mg}{2} line and 4152 km s$^{-1}$ relative to the [\ion{C}{2}] line. 
The substantial blueshift of the \ion{C}{4} broad emission line, consistent with what is expected from the relation between the outflow velocity and the \ion{C}{4} blueshift \citep{Rankine20}, is among the most extreme cases observed at lower redshifts.
Such large blueshifts of \ion{C}{4} emission have also commonly been observed in other luminous $z>6.5$ quasars \citep[e.g.][]{Meyer19, Mazzucchelli17,Wang20b}, but have not been commonly found in lower redshift quasars.
These observations indicate that the earliest SMBHs are in a fast-growing phase (i.e. with a high Eddington ratio) and that quasar outflows could play a crucial role in regulating the growth of the earliest SMBHs and their hosts.

\section{Quasar Host Galaxy} \label{sec_host}
The dust continuum around the redshifted [\ion{C}{2}] emission from the quasar host galaxy is significantly detected by ALMA (Figure \ref{fig_c2}).
We measure the continuum flux $S_{\rm 228.4~GHz}=0.45\pm0.05$ mJy from a two-dimensional Gaussian fit to the collapsed continuum map using the CASA task {\tt imfit}. The continuum emission is marginally resolved with a de-convolved size of $0\farcs62\times0\farcs29$ or a physical size of $3.12\times1.44 ~{\rm kpc}$, similar to the sizes of other quasar host galaxies at $z\gtrsim6$ \citep[e.g.][]{Venemans18}.
Under the standard assumption of an optically thin greybody dust spectral energy distribution with a dust temperature of $T_{\rm dust}=47~{\rm K}$ and an emissivity index of $\beta=1.6$ \citep{Beelen06}, after considering the effect of the cosmic microwave background on the dust emission, we estimate a total infrared luminosity of $L_{\rm TIR}=(1.5\pm0.2)\times10^{12}~L_{\odot}$. 
We estimate the star formation rate (SFR) of the quasar host galaxy to be SFR=225$\pm$25 $M_\odot~{\rm yr^{-1}}$ based on the scaling relation between the SFR  and $L_{\rm TIR}$ \citep{Murphy11}. With the same set of assumptions, we estimate a total dust mass of $\sim7\times10^7~M_\odot$ following \cite{Venemans18}. 

The [\ion{C}{2}] line is also detected (with peak pixel at $>4\sigma$) in our ALMA observations as shown in Figure \ref{fig_c2}. A single Gaussian fit to the [\ion{C}{2}] line gives $z_{\rm [CII]}=7.6423\pm0.0013$ and $\rm FWHM_{[CII]}=312\pm94 ~{\rm km~ s^{-1}}$. 
Instead of measuring the total [\ion{C}{2}] line flux from the one-D spectrum, we measure it from the two-dimensional integrated line map using {\tt imfit} following \cite{Decarli18}, and derive $F_{\rm [CII]} = 0.60\pm0.16~ {\rm Jy~ km ~s^{-1}}$. 
This corresponds to a [\ion{C}{2}] luminosity of $L_{\rm [CII]} = (0.80\pm0.22)\times 10^9~L_{\odot}$ and a SFR$_{\rm [CII]}=40-240~M_\odot ~ {\rm yr^{-1}}$ by adopting the empirical relation from \cite{DeLooze14}, consistent with the SFR derived from the total infrared luminosity. The properties of the quasar host galaxy are comparable to that of the other two quasar host galaxies known at $z>7.5$ \citep{Venemans17,Yang20a}.
Future high-resolution observations of the star formation distribution with ALMA and observations of the kinematics of extended [\ion{O}{3}] emission with the Near-Infrared Spectrograph (NIRSpec) on James Webb Space Telescope (JWST) will be necessary to carry out more detailed investigations of the dynamical mass of the host galaxy and the impact of the quasar outflow on star formation in the host galaxy.

\section{Summary} \label{sec_summ}
In this letter, we report the discovery of a luminous quasar, J0313--1806, at redshift $z=7.642$. It is the most distant quasar yet identified. J0313--1806 has a bolometric luminosity of $L_{\rm bol}=(1.4\pm0.1)\times10^{47}~{\rm erg~s^{-1}}$ and hosts a SMBH with a mass of $(1.6\pm0.4)\times10^9~M_\odot$, accreting at an Eddington ratio of $0.67\pm0.14$. The existence of such a SMBH just $\sim670$ million years after the Big Bang puts strong constraints on the formation models of seed black holes.
The quasar's rest-frame UV spectrum exhibits broad absorption troughs from extremely high-velocity outflows. These outflows have a maximum velocity up to $\sim20$\% of the speed of light. 
We also detect strong dust emission and [\ion{C}{2}] line emission from the host galaxy in ALMA data. The ALMA observations suggest that J0313--1806 is hosted by an intensely star-forming galaxy with a star formation rate of $\sim200~M_\odot~{\rm yr^{-1}}$. The continuum observations indicate that substantial dust ($\sim 7\times10^7~M_\odot$) was already built up in the quasar host galaxy. 
The relativistic quasar outflow and the fast SMBH growth phase, combined with the intense star forming activity in the host galaxy, suggest that J0313--1806 is an ideal target for investigating the assembly of the earliest SMBHs and their massive host galaxies with future high resolution ALMA and JWST NIRSpec/IFU observations. 

\acknowledgments
Support for this work was provided by NASA through the NASA Hubble Fellowship grant \#HST-HF2-51448.001-A awarded by the Space Telescope Science Institute, which is operated by the Association of Universities for Research in Astronomy, Incorporated, under NASA contract NAS5-26555.
J.Y., X.F., and M.Y. acknowledge the support from the US NSF grant AST 15-15115, AST 19-08284, and NASA ADAP grant NNX17AF28G. Research by A.J.B. is supported by NSF grant AST-1907290. The work of T.C. was carried out at the Jet Propulsion Laboratory, California Institute of Technology, under a contract with NASA. 
ACE acknowledges support by NASA through the NASA Hubble Fellowship grant $\#$HF2-51434 awarded by the Space Telescope Science Institute, which is operated by the Association of Universities for Research in Astronomy, Inc., for NASA, under contract NAS5-26555.
L.J. acknowledge support from the National Science Foundation of China (11721303, 11890693) and the National Key R\&D Program of China (2016YFA0400703). B.P.V. acknowledges funding through the ERC Advanced Grant 740246 (Cosmic Gas). X.W. is thankful for the support from the National Key R\&D Program of China (2016YFA0400703) and the National Science Foundation of China (11533001 \& 11721303).

We thank John Blakeslee and Sean Dougherty for approving the Director's Discretionary Time observation requests for Gemini/GNIRS and ALMA, respectively.  
We thank Rene Rutten and Hwihyun Kim for their supports on the Gemini/Flamingos observations.
Thanks to Dave Osip for approving the request of Magellan/FIRE spectroscopy, which is important to the confirmation of this quasar.
We thank the anonymous referee for carefully reading the manuscript and providing comments.

This paper makes use of the following ALMA data: ADS/JAO.ALMA\#2019.A.00017.S. ALMA is a partnership of ESO (representing its member states), NSF (USA) and NINS (Japan), together with NRC (Canada), MOST and ASIAA (Taiwan), and KASI (Republic of Korea), in cooperation with the Republic of Chile. The Joint ALMA Observatory is operated by ESO, AUI/NRAO and NAOJ. In addition, publications from NA authors must include the standard NRAO acknowledgement: The National Radio Astronomy Observatory is a facility of the National Science Foundation operated under cooperative agreement by Associated Universities, Inc.

This research is based in part on observations obtained at the international Gemini Observatory (GS-2019B-Q-134, GN-2019B-DD-110), a program of NSF’s NOIRLab, which is managed by the Association of Universities for Research in Astronomy (AURA) under a cooperative agreement with the National Science Foundation. on behalf of the Gemini Observatory partnership: the National Science Foundation (United States), National Research Council (Canada), Agencia Nacional de Investigaci\'{o}n y Desarrollo (Chile), Ministerio de Ciencia, Tecnolog\'{i}a e Innovaci\'{o}n (Argentina), Minist\'{e}rio da Ci\^{e}ncia, Tecnologia, Inova\c{c}\~{o}es e Comunica\c{c}\~{o}es (Brazil), and Korea Astronomy and Space Science Institute (Republic of Korea).

Some of the data presented in this paper were obtained at the W. M. Keck Observatory, which is operated as a scientific partnership among the California Institute of Technology, the University of California, and the National Aeronautics and Space Administration. The Observatory was made possible by the generous financial support of the W. M. Keck Foundation. The authors wish to recognize and acknowledge the very significant cultural role and reverence that the summit of Maunakea has always had within the indigenous Hawaiian community. 

This paper includes data gathered with the 6.5 meter Magellan Telescopes located at Las Campanas Observatory, Chile.

\vspace{5mm}
\facilities{ALMA, Magellan (FIRE), Gemini (FLAMINGOS-2), Gemini (GNIRS), Keck (NIRES)}
\software{PypeIt \citep{pypeit1,pypeit2}, CASA \citep{CASA}}

\end{document}